\documentclass[11pt,preprintnumbers,aps,amssymb,nofootinbib,amsmath]
{revtex4}
\usepackage{epsfig,epsf}
\usepackage{bm} % puts greek and math symbols in boldface using \bm
\newcommand{\beq}{\begin{equation}}
\newcommand{\beql}[1]{\begin{equation}\label{#1}}
\newcommand{\eeq}{\end{equation}}
\newcommand{\bea}{\begin{eqnarray}}
\newcommand{\eea}{\end{eqnarray}}
\newcommand{\eq}[1]{(\ref{#1})}
\newcommand{\fig}[1]{Fig.~\ref{#1}}
\renewcommand{\sec}[1]{Sec.~\ref{#1}}
\newcounter{topiccounter}
\renewcommand{\b}[1]{\mathbf{#1}}
\newcommand{\unit}[1]{\hat {\mathbf{#1}}} % unit vector

\newcommand{\as}{\alpha_s}
\newcommand{\bas}{\bar\alpha_s}

\newcommand{\real}{\mathrm{Re}\,}

\begin{document}

\preprint{RBRC-825}

\title{Rapidity and centrality dependence of azimuthal correlations in Deuteron-Gold collisions at RHIC}

\author{Kirill Tuchin$\,^{a,b}$\\}

\affiliation{
$^a\,$Department of Physics and Astronomy, Iowa State University, Ames, IA 50011\\
$^b\,$RIKEN BNL Research Center, Upton, NY 11973-5000\\}

\date{\today}


\begin{abstract}

We  calculate azimuthal correlations  in $dAu$ collisions at different rapidities and centralities  and argue that experimentally observed depletion of the back-to-back pick can be quantitatively explained by gluon saturation  in the Color Glass Condensate of the Gold nucleus.  

\end{abstract}

\maketitle

%%%%%%%%%%%%%%%%%%%%%%%%%%%%%%%%%%%%%%%%
\section{Introduction}\label{sec:intr}

The subject of this paper is quantitative investigation of azimuthal two-particle correlations in deuteron ($d$) -- Gold ($Au$) collisions at RHIC. These correlations are an important tool in studying  properties of the Color Glass Condensate (CGC) \cite{GLR,MUQI,MV,JIMWLK1,JIMWLK2} -- a coherent quasi-classical state of fast gluons and quarks discovered at RHIC \cite{Gyulassy:2004zy}. In heavy ion collisions they are the main source of azimuthal anisotropy at semi-hard transverse momenta \cite{Kovchegov:2002nf,Kovchegov:2002cd,Tang:2004vc}. While the azimuthal correlations in heavy-ion collisions are smeared by the collective `flow' effects, those in deuteron--heavy ion ($dA$) collisions allow for the direct observation of the CGC effect on the correlation function \cite{Kharzeev:2004bw}.  A qualitative analysis of \cite{Kharzeev:2004bw} predicted most of the features of the azimuthal correlation function. However, a more precise data that recently has become available urges for a more quantitative approach, which we undertake in this article. 

The general form of the back-to-back correlations in hard processes in the kinematic region of large $x$ follows from momentum conservation in the leading $2\to2$ process. In the center-of-mass frame the outgoing partons move in opposite directions with the same momenta implying a sharp pick at the azimuthal opening angle of $\Delta\phi=\pi$ independently of rapidity and centrality. It is essential for this argument that incoming partons are approximately on-mass-shell. This assumption breaks down when the produced particle transverse momenta become comparable with the characteristic transverse momentum scale of fast hadron or nucleus. At small $x$ this scale is called the saturation scale $Q_s$ \cite{GLR}  and is known to increase with centrality and decrease with $x$ \cite{GLR,MUQI,Kovchegov:1998bi,Levin:1999mw}, see  \eq{satsc}. At the central rapidity at RHIC, its numerical value for a Gold nucleus is $Q_s\simeq 1.5$~GeV. Therefore, production of particles with comparable or smaller transverse momenta is strongly affected by CGC. Since $x$ of nucleus decreases towards the deuteron fragmentation region, one expects even stronger effect of CGC on particle  production in the forward direction \cite{Kharzeev:2003wz,Kharzeev:2004yx,Albacete:2003iq}. Since the incoming partons are off-mass-shell, $2\to 3$ process becomes the leading one in particle production yielding the uncorrelated background. This is natural, because CGC is a classical field in which correlations are suppressed.  Correlations appear as quantum fluctuations around the classical field and correspond to $2\to 4$ processes. The corresponding azimuthal correlation function also  picks at  $\Delta \phi =\pi$, which, however,  is not as sharp as in the hard processes since the  momentum conservation no longer  requires the outgoing partons to be back-to-back. This is the main source for depletion of back-to-back correlations for particles separated by small rapidity interval $\Delta y\ll 1$. At large rapidity separations, the rapidity gap becomes filled with semi-hard partons further depleting  the correlation.

At present there are two heuristic approaches to take the CGC effects into account for observable quantities of interest.  One is based on the `dipole model' \cite{dipole} in which one reduces the relevant scattering amplitudes  to a product of light-cone ``wave-functions" and combinations of CGC field correlators in the configuration space. These correlators satisfy a set of evolution equations with certain initial conditions. The advantage of this approach is that it rests on accurate  theoretical treatment of the gluon saturation region. In this approach the double inclusive gluon \cite{JalilianMarian:2004da}, quark--anti-quark \cite{Tuchin:2004rb,Blaizot:2004wv,Kovchegov:2006qn}   and valence quark -- gluon  \cite{Marquet:2007vb} cross sections were calculated assuming large rapidity gap $\Delta y$ between the produced pair of partons (multi-Regge kinematics, MRK). It was argued in \cite{Leonidov:1999nc,Kovchegov:2002cd} that this approximation misses important features of azimuthal angle dependence at not too large $\Delta y$. At the moment, extending the dipole model to the region of $\Delta y\sim 1$ presents a serious computational challenge.

 Another approach is `$k_T$-factorization', which assumes that $2\to n$ process and  the two-point correlation functions of CGC fields  can be factored out. In this approximation,  the $2\to 4$ amplitudes were calculated for an arbitrary $\Delta y$ (quasi multi-Regge kinematics, QMRK) in \cite{LRSS,CCH,CE,Fadin:1997hr} for $gg\to ggq\bar q$ and in \cite{Fadin:1996zv,Leonidov:1999nc,Bartels:2006hg} for $gg\to gggg$ processes. 
 The corresponding result is given by  \eq{Atotal}-\eq{ggqq}. Although generally  $k_T$-factorization fails in the gluon saturation region, there are valid reasons to believe that it provides a \emph{reasonable approximation} of the observed quantities. Indeed, it was proved that $k_T$-factorization provides the exact result for the cross section for single inclusive gluon production in the leading logarithmic approximation (LLA) \eq{sincl} \cite{Kovchegov:2001sc} (though there is  a subtlety in the definition of the unintegrated gluon distribution $\varphi$ \cite{Kovchegov:2001sc,Kharzeev:2003wz}). Although $k_T$-factorization fails for the double-inclusive heavy quark production,  the deviation from the exact results is not large at RHIC energies  \cite{Fujii:2005vj}. At  transverse momenta of produced particles much larger than $Q_s$, $k_T$-factorization rapidly converges to the exact results. There are also numerous indications that  $k_T$-factorization is  phenomenologically reliable. Thus, the KLN model \cite{Kharzeev:2000ph,Kharzeev:2001yq,Kharzeev:2001gp,Kharzeev:2002ei}  that relies on $k_T$-factorization provides successful description of experimental data. Recently, long range rapidity correlations in heavy-ion collisions were calculated in \cite{Dumitru:2008wn,Gelis:2008ad,Gelis:2008sz,Dusling:2009ni}. Although their theoretical results are more general, the phenomenological applications assume $k_T$-factorization. In view of these arguments, we will use $k_T$-factorization to compute the azimuthal correlations in this work.     

The paper is structured as follows.  In \sec{sec2}  we compute the azimuthal correlation function at small rapidity separations $\Delta y\ll 1$ at midrapidity $y=0$ and forward rapidity $y=3$. In \sec{sec3} we consider azimuthal correlations of particles separated by large rapidity gap $\Delta y=3$. We discuss a possible effect of gluon evolution in the gap. Our calculations in both sections agree reasonably well with experimental data and provide an additional support for the CGC description of the nuclear wave function. 
We conclude in \sec{sec:concl}.

%%%%%%%
\section{Correlations at $|y_T-y_A|\lesssim 1$ }\label{sec2}

Azimuthal correlation function is defined as 
\beq\label{cf}
C(\Delta \phi)=\frac{1}{N_{\mathrm{trig}}}\frac{dN}{d(\Delta\phi)}\,,
\eeq
where $dN/d(\Delta\phi)$ is the number of pairs produced in the given opening angle $\Delta\phi$ and $N_{\mathrm{trig}}$ is the number of trigger particles. 
The number of pairs is given by
\beq\label{npa}
\frac{dN}{d(\Delta\phi)}= 2\pi \int dk_T k_T \int d y_T\,\int  dk_A k_A \int dy_A
\left( \frac{dN_\mathrm{trig}}{d^2k_Tdy_T}  \frac{dN_\mathrm{ass}}{d^2k_Ady_A}
+ \frac{dN_\mathrm{corr}}{d^2k_Tdy_T\,d^2k_Ady_A}  \right)
\eeq
where $\b k_T$ and $y_T$ are the transverse momentum and rapidity  of the trigger particle and   $\b k_A$ and $y_A$ are the transverse momentum and rapidity  of the associate one. We denote $k_T=\sqrt{\b k_T^2}$ etc.\ throughout this paper.
The first term on the r.h.s.\ of \eq{npa} corresponds to gluon production in two different  sub-collisions (i.e.\ at different impact parameters) and therefore gives a constant contribution to the correlation function, whereas the second term on the r.h.s.\  describes production of two particles in the same sub-collision.  The number of the trigger particles is given by 
\beq\label{strig}
N_\mathrm{trig}= 2\pi \int dk_T k_T\int dy_T\, \frac{dN_\mathrm{trig}}{d^2k_Tdy_T}\,.
\eeq
Expression for the single inclusive  gluon cross section is well-known \cite{Levin:1974fw,Laenen:1994gh,Kovchegov:1997ke,Gyulassy:1997vt,Kovchegov:1998bi,Braun:2000ca,Braun:2001kh,Kovchegov:2001sc,Blaizot:2004wu}. The corresponding multiplicity reads
 \beql{sincl}
 \frac{dN}{d^2 k\,dy}\,=\,
\frac{2 \, \as\,}{C_F \, S_\bot}\,\frac{1}{ k^2}\,\int\,
d^2 q_1\varphi_D(x_+,q^2_1)\,\varphi_A(x_-,(\b k- \b q_1)^2)\,,
\eeq
where the unintegrated gluon distribution function $\varphi$ is simply related to the gluon distribution function as 
\beql{gdf}
xG(x,Q^2)= \int^{Q^2}dq^2 \varphi(x,q^2)\,.
\eeq
$x_\pm$ are the fractions of momenta of incoming nucleons carried away by the $t$-channel gluons; in the center-of-mass frame
\beql{xx}
x_\pm=\frac{k}{\sqrt{s}}\, e^{\pm y}\,.
\eeq
Equation \eq{sincl} is derived in the multi-Regge kinematics (MRK) $x_\pm\ll 1$. 

The correlated part of  double-inclusive parton multiplicity is given by 
\begin{eqnarray}
\frac{dN_\mathrm{corr}}{d^2 k_T\,dy_T\, d^2 k_A\,dy_A}&=&
\frac{N_c \, \as^2}{\pi^2 \, C_F \, S_\bot}\,
\int\,
\frac{d^2 q_1}{q_1^2}\,\int\,\frac{d^2q_2}{q_2^2}\,\delta^2(\b q_1+\b q_2-\b k_T-\b k_A)\nonumber\\
&&\times\,\varphi_D(x_1,q^2_1)\,\varphi_A(x_2,q^2_2)\,\mathcal{A}(\b q_1,\b q_2, \b k_T, \b k_A,y_T-y_A)\,,\label{nlow}
\end{eqnarray}
where 
\beql{x12}
x_1=\frac{k_{T}e^{y_T} + k_{A}e^{y_A}}{\sqrt{s}}\,,\qquad
x_2=\frac{k_{T}e^{-y_T} + k_{A}e^{-y_A}}{\sqrt {s}}\,.
\eeq
The amplitude $\mathcal{A}$ was computed in the quasi-multi-Regge-kinematics (QMRK) in \cite{Fadin:1996zv,Fadin:1997hr,Leonidov:1999nc} and recently re-derived in \cite{Bartels:2006hg} (the $gg\to ggq\bar q $ part was calculated before in \cite{LRSS,CCH,CE}). In QMRK one assumes that $x_1,x_2\ll 1$, but $\Delta y $ is finite. 
To write down the amplitude $\mathcal{A}$ it is convenient to introduce the kinematic invariants in a usual way. We will also denote $\b k_T=\b k_1$, $\b k_A=\b k_2$ and $y_T- y_A= \Delta y$ to simplify notations. We have \cite{Leonidov:1999nc}
\begin{subequations}
\begin{eqnarray}
\hat s&=&2(k_1k_2\cosh(\Delta y)-\b k_{1}\cdot \b k_{2})\,,\\
\hat t&=&-(\b q_{1}-\b k_{1})^2-k_1k_2e^{\Delta y}\,,\\
\hat u&=&-(\b q_{1}-\b k_{2})^2-k_1k_2e^{-\Delta y}\,,\\
\Sigma&=&x_1x_2s=k_1^2+k_2^2+2k_1k_2\cosh(\Delta y)\,.
\end{eqnarray}
\end{subequations}
Now the amplitude $\mathcal{A}$ reads
\beq\label{Atotal}
{\cal A}={\cal A}_{gg\to gg}+\frac{n_f}{4N^3_c}{\cal A}_{gg\to q\bar q}\,,
\eeq
where
\begin{eqnarray}
{\cal A}_{gg\to gg}=
&q_1^2q_2^2&
    \left\{ -\frac{1}{\hat t\hat u}+\frac{1}{4\hat t\hat u}\frac{q_1^2q_2^2}{k_1^2k_2^2}-
            \frac{e^{\Delta y}}{4\hat tk_1k_2}-\frac{e^{-\Delta y}}{4\hat uk_1k_2}+
            \frac{1}{4k_1^2k_2^2}+
    \right.\nonumber\\
&&         \frac{1}{\Sigma}
             \left[-\frac{2}{\hat s}
                \left(1+k_1k_2(\frac{1}{\hat t}-\frac{1}{\hat u})\sinh(\Delta y)
                \right)+
                \frac{1}{2k_1k_2}(1+\frac{\Sigma}{\hat s})\cosh(\Delta y)-
             \right.\nonumber\\
&&
               -\frac{q_1^2}{4\hat s}
                      [(1+\frac{k_2}{k_1}e^{-\Delta y})\frac{1}{\hat t}+
                       (1+\frac{k_1}{k_2}e^{\Delta y})\frac{1}{\hat u}]\nonumber\\
&&           \left.
    \left.
               -\frac{q_2^2}{4\hat s}
                      [(1+\frac{k_1}{k_2}e^{-\Delta y})\frac{1}{\hat t}+
                       (1+\frac{k_2}{k_1}e^{\Delta y})\frac{1}{\hat u}]
             \right]
    \right\}\nonumber \\
 &&   + \frac{1}{2}\left\{\left(
\frac{(\b k_{1}-\b q_{1})^2
(\b k_{2}-\b q_{1})^2-k_1^2k_2^2}{\hat t\hat u}\right)^2
\right.\nonumber\\
&&-\frac{1}{4}
\left.\left( \frac{(\b k_{2}-\b q_{1})^2-k_1k_2e^{-\Delta y}}
                  {(\b k_{2}-\b q_{1})^2+k_1k_2e^{-\Delta y}}
             -\frac{E}{\hat s}
      \right)
      \left( \frac{(\b k_{1}-\b q_{1})^2-k_1k_2e^{\Delta y}}
                  {(\b k_{1}-\b q_{1})^2+k_1k_2e^{\Delta y}}
             +\frac{E}{\hat s}
      \right)
\right\}\,,
\label{gggg}
\end{eqnarray}
with 
\beq
E=(\b q_{1}-\b q_{2})\cdot (\b k_{1}-\b k_{2})-\frac{1}{\Sigma}
(q_1^2-q_2^2)(k_1^2-k_2^2)+2k_1k_2\sinh(\Delta y)
\left(1-\frac{q_1^2+q_2^2}{\Sigma}\right)\,,
\eeq
and
\begin{eqnarray}
{\cal A}_{gg\to q\bar q}&=&N_c^2\,\left\{ 2\frac{q_{1}^{2}q_{2}^{2}}{\hat s\Sigma }
\left(1+k_{1}k_{2}\sinh (\Delta y)(\frac{1}{\hat t}-\frac{1}{\hat u})\right)
-\left( \frac{(\b k_{1}-\b q_{1})^{2}(\b k_{2}-\b q_{1})^{2}
-k_{1}^{2}k_{2}^{2}}{\hat t\hat u}\right)^{2}\right.   \nonumber \\
&&+\left. \frac{1}{2}
      \left( \frac{(\b k_{2}-\b q_{1})^2-k_1k_2e^{-\Delta y}}
                  {(\b k_{2}-\b q_{1})^2+k_1k_2e^{-\Delta y}}
             -\frac{E}{\hat s}
      \right)
      \left( \frac{(\b k_{1}-\b q_{1})^2-k_1k_2e^{\Delta y}}
                  {(\b k_{1}-\b q_{1})^2+k_1k_2e^{\Delta y}}
             +\frac{E}{\hat s}
      \right)
\right\}\nonumber\\
&& +  \left( \frac{(\b k_{1}-\b q_{1})^{2}
(\b k_{2}-\b q_{1})^{2}-k_{1}^{2}k_{2}^{2}}{\hat t\hat u}\right)^{2}
-\frac{q_{1}^{2}q_{2}^{2}}{\hat t\hat u}\,.
\label{ggqq}
\end{eqnarray}
In the Leading Logarithmic Approximation  $\Delta y\sim 1/\as\gg 1$  the amplitude factorizes (Multi-Regge Kinematics) becoming
\beq\label{mrk}
{\cal A}={\cal A}_{gg\to gg}=\frac{q_1^2q_2^2}{k_1^2k_2^2}\,,\quad \Delta y\to \infty\,.
\eeq
The double-inclusive cross section in MRK  was  first calculated in \cite{Levin:1974fw}.

For numerical calculations we need a model for the unintegrated gluon distribution function $\varphi$.  In spirit of the KLN model \cite{Kharzeev:2000ph,Kharzeev:2001yq,Kharzeev:2001gp} we write 
\beq\label{kln}
\varphi(x,q^2)=\frac{1}{2\pi^2}\frac{S_\bot C_F}{\as}\big(1-e^{-Q_s^2/q^2}\big) \, (1-x)^4\,.
\eeq 
At $e^{-1/\as} \ll x\ll 1$ and $q^2\gg Q_s^2$ \eq{kln} and \eq{gdf} yield  $xG=A\frac{\as C_F}{\pi}\ln Q^2$ as required. The saturation scale is given by
\beql{satsc}
Q_s^2= A^{1/3} \bigg(\frac{x_0}{x}\bigg)^\lambda\,,
\eeq
where $x_0=3.04\cdot 10^{-4}$ and $\lambda=0.288$ are fixed by fits of the DIS data \cite{GolecBiernat:1999qd,GolecBiernat:1998js}. The coupling constant is fixed at $\as=0.3$. 

It has been pointed out in \cite{Leonidov:1999nc} that due to $1\to 2$ gluon splittings the double-inclusive cross section has a collinear  singularity at $\hat s\to 0$, i.e.\ it is proportional to $[(\Delta y)^2  +(\Delta \phi)^2]^{-1}$. Such singularities are usually cured at the higher orders of the perturbation theory. Additional contributions to the small angle correlations arise from various soft processes including resonance decays, hadronization, HBT correlations etc. Because the small angle correlations are beyond the focus of the present paper we simply regulate it by imposing a cutoff on the  minimal possible value of the invariant mass $\hat s$. This is done by redefining the amplitude as $\mathcal{A}\to \mathcal {A} \,\hat s/(\mu^2+\hat s)$. For each kinematic region, parameter $\mu$ is fixed in such a way as to reproduce the value of the correlation function in $pp$ collisions  at zero opening angle $\Delta\phi=0$. 

Expressions for the single and double inclusive cross sections written in this  section are derived in the framework of  $k_T$-factorization. As mentioned in Introduction, at the LLA this factorization gives the correct result for the single inclusive cross section. However, it fails for the double-inclusive cross section. The double inclusive gluon production was calculated in \cite{JalilianMarian:2004da} and the double-inclusive quark-antiquark production in \cite{Tuchin:2004rb,Blaizot:2004wv,Kovchegov:2006qn} both in the LLA. Unfortunately, the LLA misses many important details of the azimuthal angle dependence. On the other hand,  $k_T$-factorization is known to give results that are in qualitative agreement with a more accurate approaches, but misses the overall normalization. Therefore, in order to correct the overall normalization of the cross sections we multiply the single inclusive cross section \eq{sincl} by a constant $K_1$ and the double-inclusive one \eq{nlow} by a different constant  $K_2$ \cite{Kovchegov:2002nf}. The correlation function $C$ depends on both $K_1$ and $K_2$. However, the difference $C_\Delta=C(\Delta\phi)-C(\Delta\phi_0)$ depends only on the ratio $K_2/K_1$. We choose $\Delta\phi_0$ in such a way  that $C(\Delta\phi_0)$ is the minimum of the correlation function. This is  analogous to the experimental procedure of removing the pedestal \cite{Adams:2003im}. The overall normalization of  the correlation function $K_2/K_1$ -- which is the only essential free parameter of our model -- is fixed to reproduce the height of the correlation function in $pp$ collisions. 

Integrations over  transverse momenta of the trigger and associated particles can be simplified if we recall that because the transverse momentum spectra of both single-inclusive and double-inclusive cross sections are very steep, integration over the range of  momenta is approximately proportional to the value of the integrand at the smallest value of momentum. Namely, if $f(k)\propto 1/k^n$ with $n\gg 1$, we have $\int_{k_\mathrm{min}}^{k_\mathrm{max}} dk \, f(k)\approx f(k_\mathrm{min})\,k_\mathrm{min}\,/(n-1)$. We absorb the unknown constant  $(n-1)$ in the overall normalization of $C_\Delta$. This is the approximation that we use in our calculations. 

\begin{figure}[ht]
\begin{tabular}{cc}
      \includegraphics[height=4.5cm]{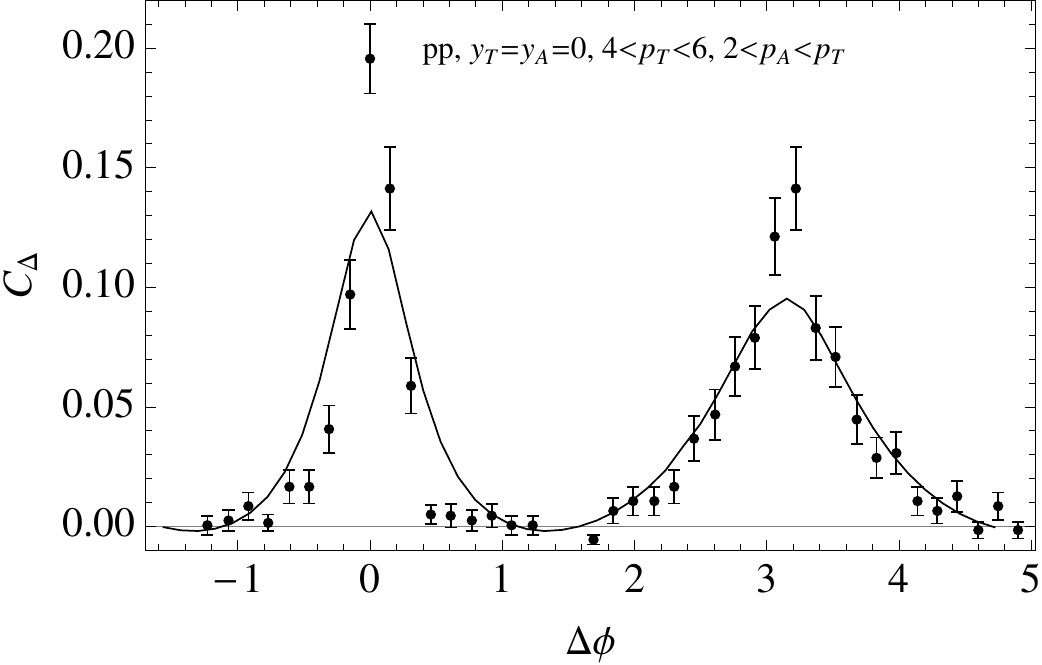} &
      \includegraphics[height=4.5cm]{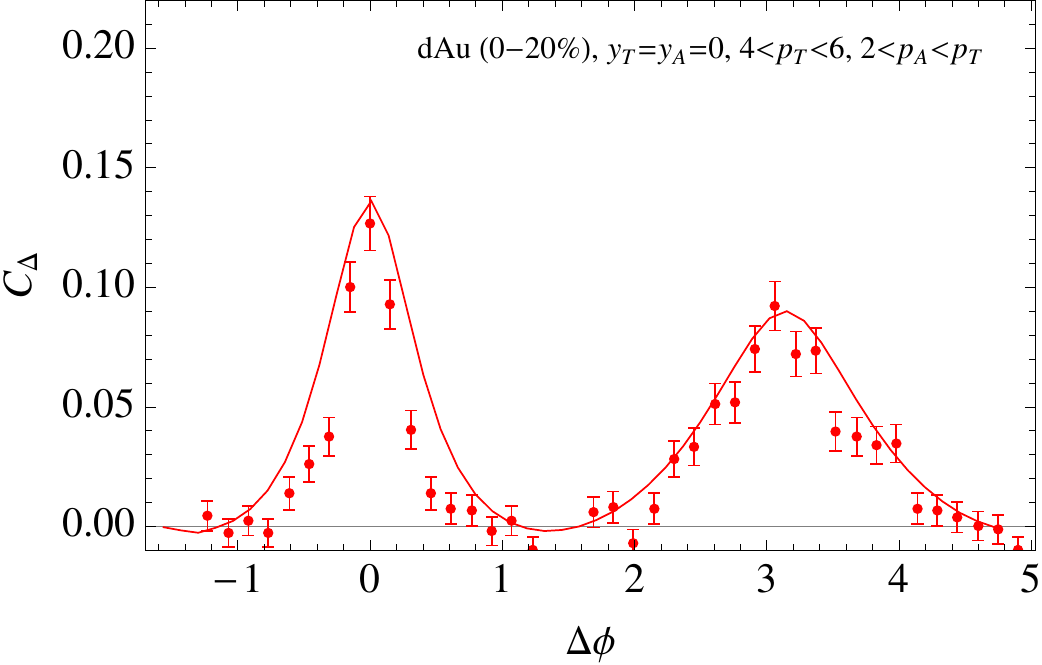} 
  \end{tabular}
  \caption{Correlation function at the central rapidity. Kinematic region is  $4<p_T <6$, $2<p_A <p_T$ (all momenta are in GeV), $y_T=3.1$, $y_A=3$. Left (right) panel: minbias $pp$ ($dAu$) collisions. Data from \cite{Adams:2003im}. }
\label{fig:cc}
\end{figure}

\begin{figure}[ht]
\begin{tabular}{cc}
      \includegraphics[height=4.5cm]{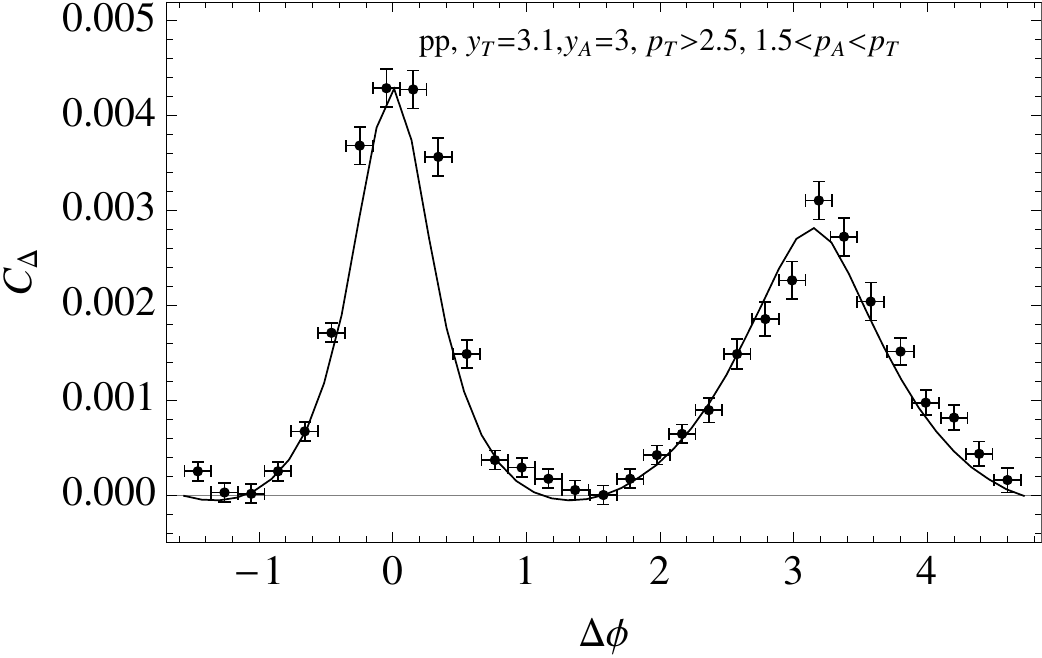} &
      \includegraphics[height=4.5cm]{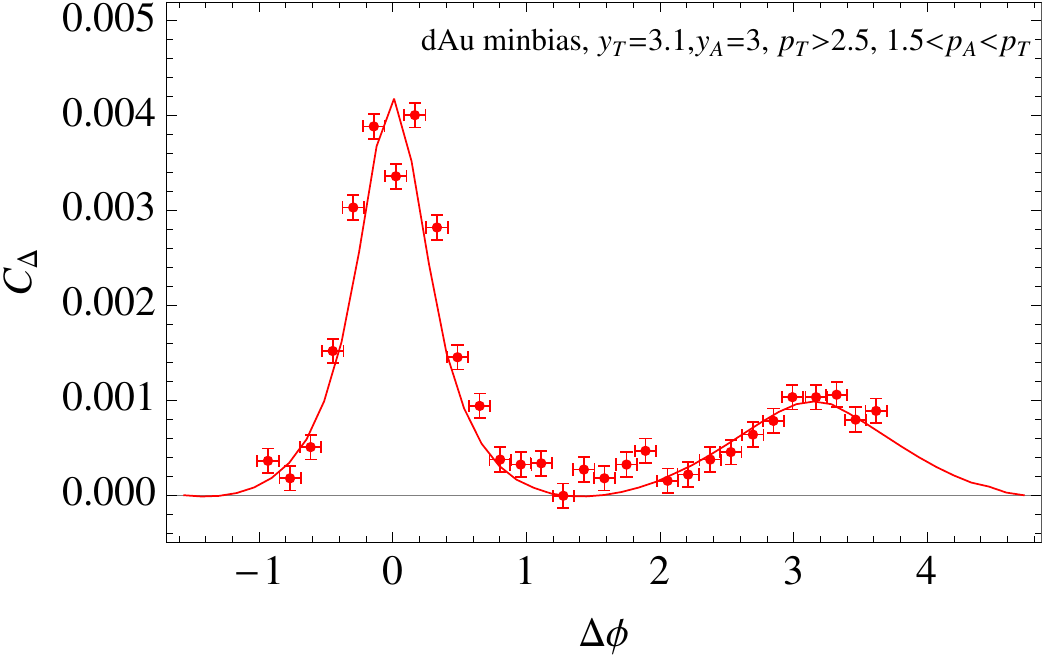} 
  \end{tabular}
  \caption{Correlation function at forward rapidities. Kinematic region is  $p_T>4$, $1.5<p_A <p_T$ (all momenta are in GeV), $y_T=3.1$, $y_A=3$. Left (right) panel:  the minbias $pp$ ($dAu$) collisions. Data from \cite{AGordon}. }
\label{fig:ff1}
\end{figure}

\begin{figure}[ht]
\begin{tabular}{cc}
      \includegraphics[height=4.5cm]{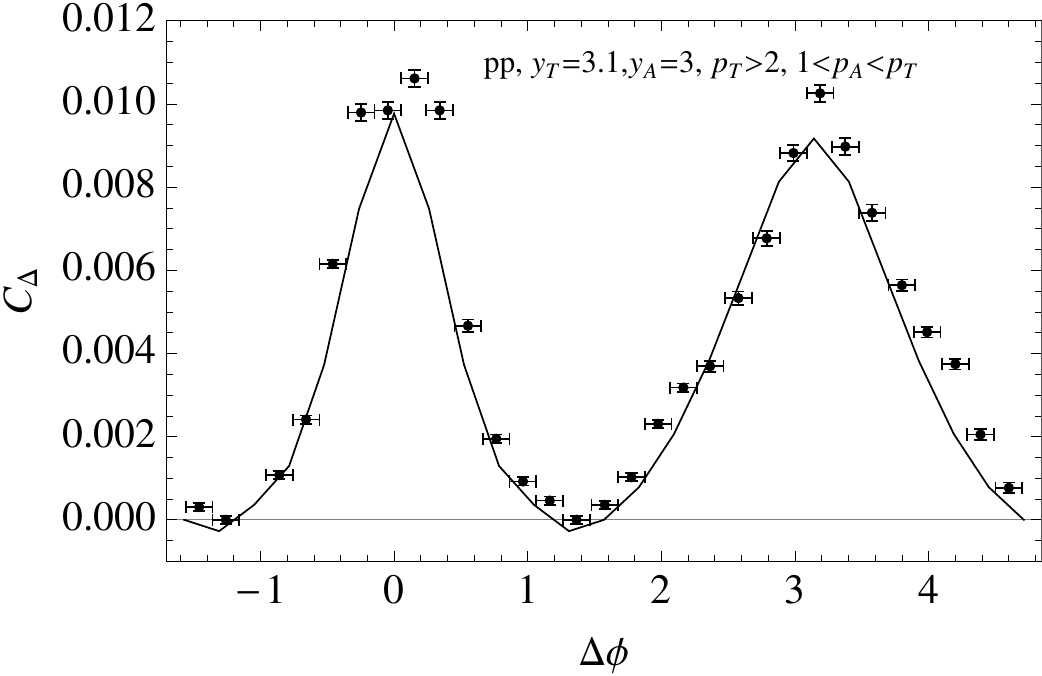} &
      \includegraphics[height=4.5cm]{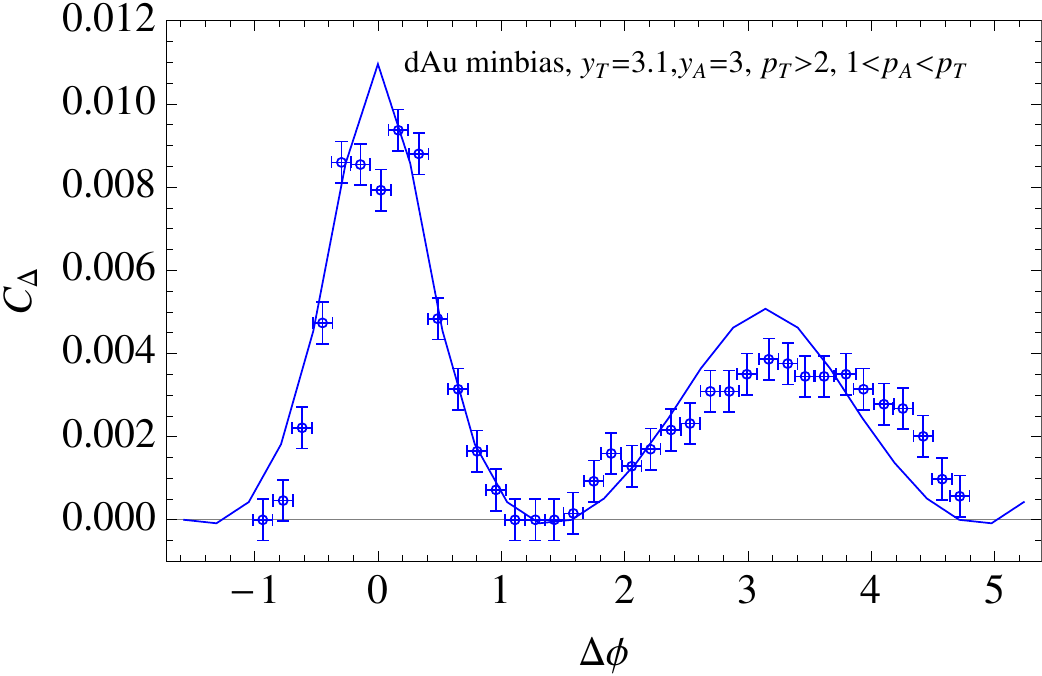} \\
      \includegraphics[height=4.5cm]{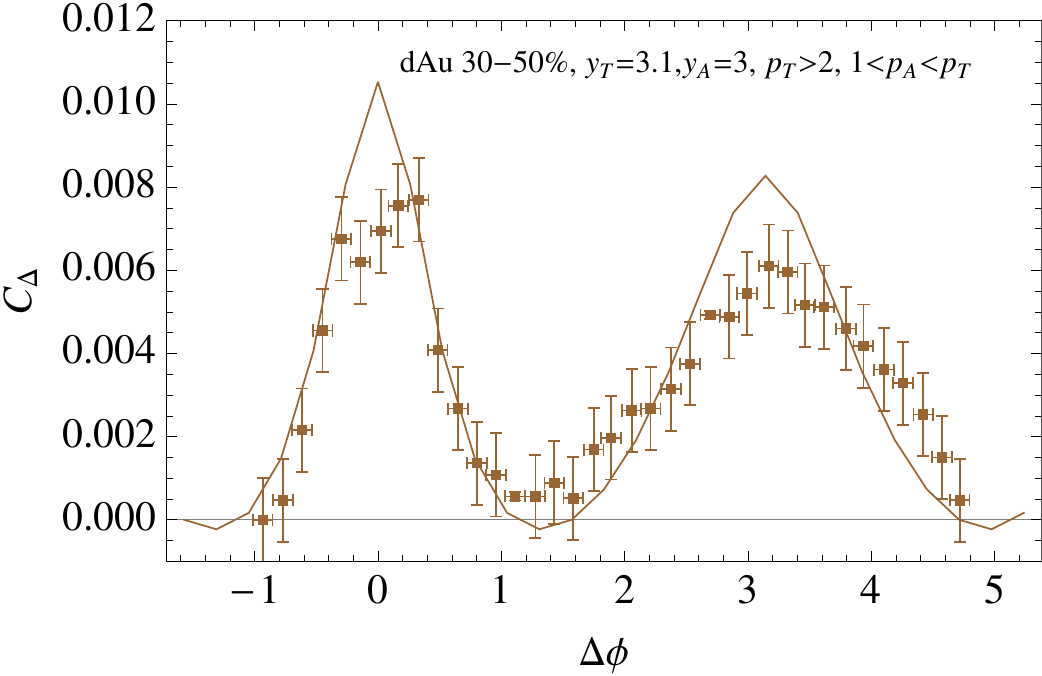}&
      \includegraphics[height=4.5cm]{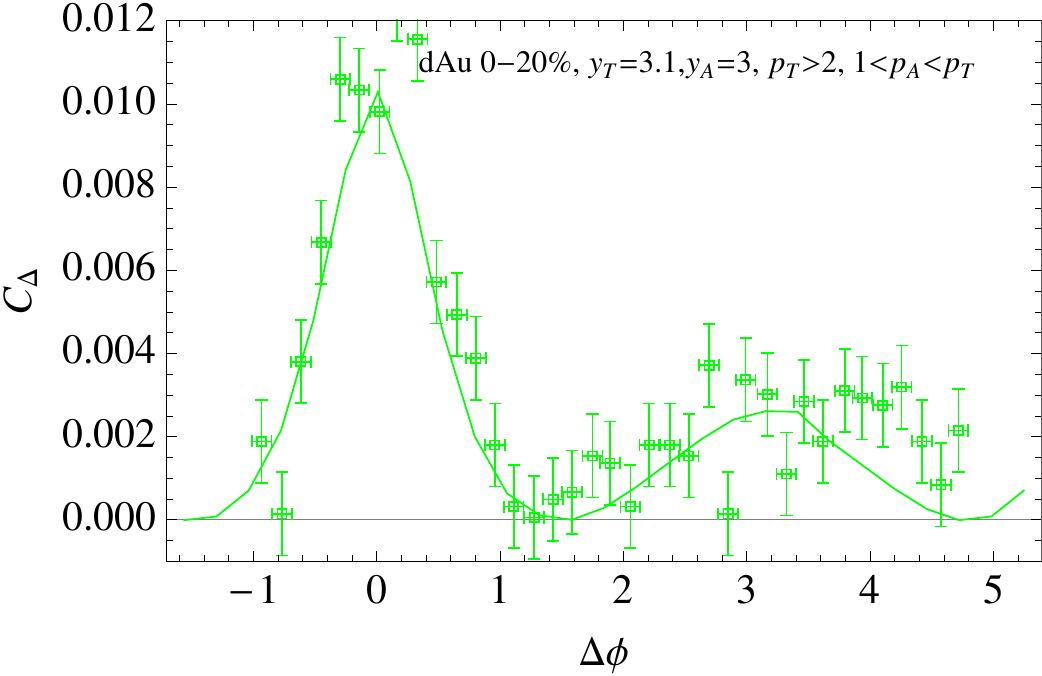}
  \end{tabular}
  \caption{Correlation function at forward rapidities. Kinematic region is  $p_T>2$, $1.5<p_A <p_T$ (all momenta are in GeV), $y_T=3.1$, $y_A=3$. Upper left (right) panel: minbias $pp$  ($dAu$) collisions. Lower left (right) panel: peripheral (central)  $dAu$ collisions. Note: centrality of the theoretical calculation may not coincide with the centrality of the data (the former is not yet known at the time of publication).   Data from \cite{AGordon}.}
\label{fig:ff2}
\end{figure}

The results of the numerical calculations are shown in \fig{fig:cc},\fig{fig:ff1} and \fig{fig:ff2}.  In these figures we observe suppression of the bak-to-back correlation in $dAu$ as compared to the $pp$ ones in agreement with the  experimental data. In \fig{fig:ff2} we also see the depletion of the  back-to-back correlation as a function of centrality. Note, that at the time of publication the precise centrality classes of the \emph{data} shown in the lower row of \fig{fig:ff2} were not known.     

%%%%
%%%%%

%%%%
%%%%%

%%%%
%%%%%

In addition to  $gg\to gggg$ and $gg\to ggq\bar q$ processes that we took into account in this section, production of valence quark of deuteron $gq_v\to gq_v gg$ gives a sizable contribution at forward rapidities due to not very small value of $x$ associated with deuteron ($x\approx 0.2$ for $p_T=2$~GeV at $y=3$). Contribution of this process to azimuthal correlations was analyzed in \cite{Nikolaev:2005dd,Marquet:2007vb} in the framework of the dipole model in MRK. However, the corresponding expression  in $k_T$-factorization in QMRK is presently unknown thus preventing us from taking it into account in our calculation.\footnote{Authors of  \cite{JalilianMarian:2004da,Marquet:2007vb}  discussed the process $gq_v\to ggq_v$  assuming the collinear approximation to the unintegrated valence quark distribution \cite{Itakura:2003jp}.}  In-spite of this we believe that the general structure of the correlation function as well as its centrality dependence is not strongly affected by the valence quark contribution. We plan to address this problem elsewhere. 

\begin{figure}[ht]
\begin{tabular}{cc}
      \includegraphics[height=4.5cm]{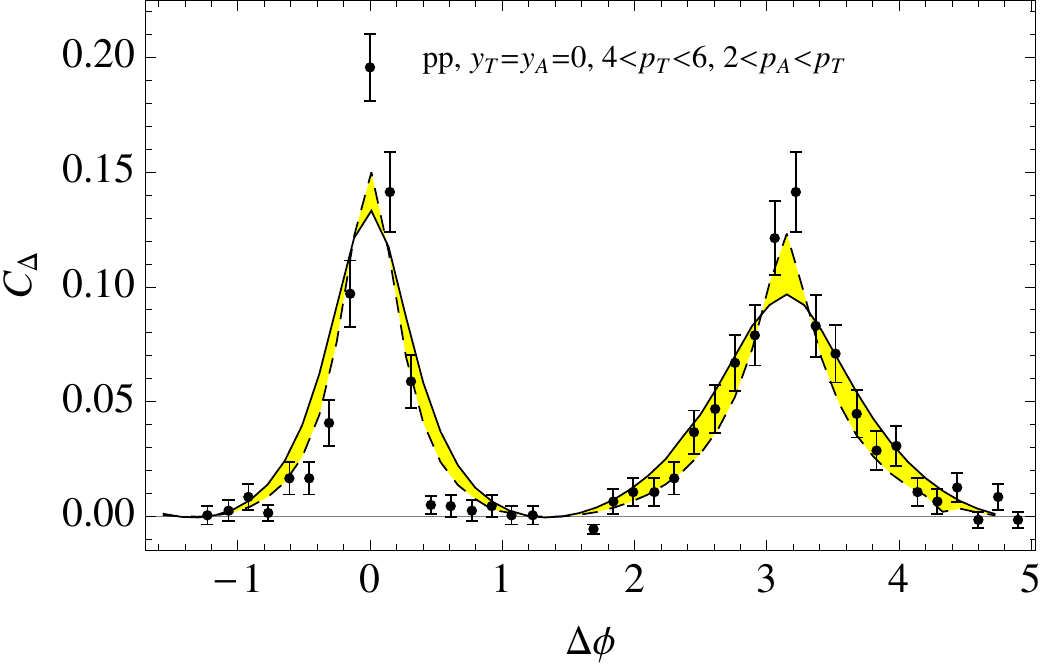} &
      \includegraphics[height=4.5cm]{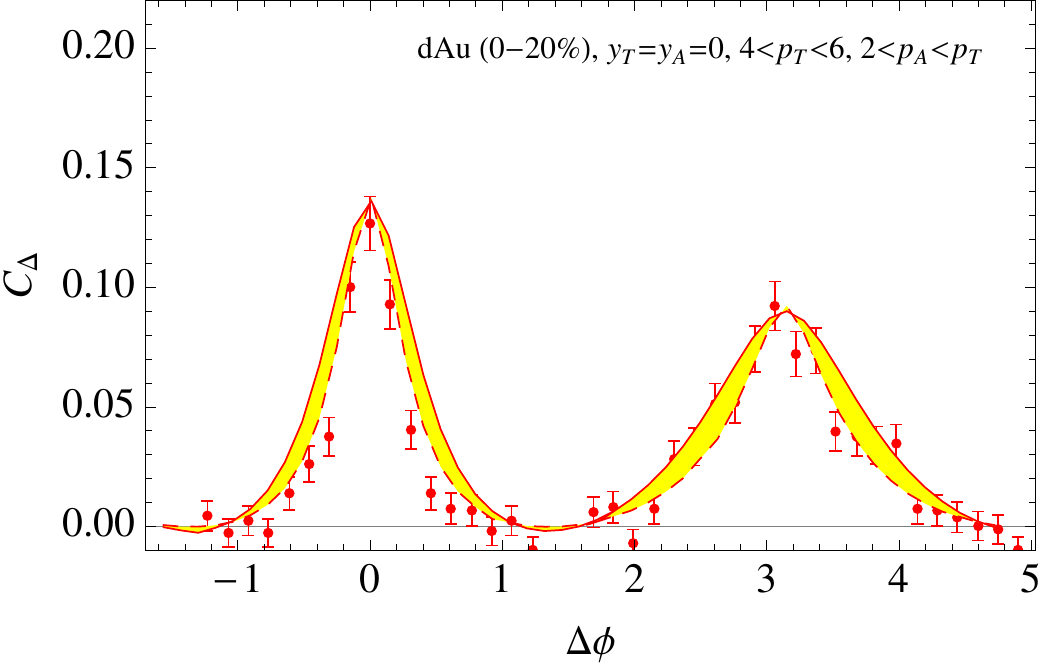} 
  \end{tabular}
  \caption{Effect of fragmentation on the azimuthal correlation function.   Solid lines are the same as in  \fig{fig:cc}. Dashed lines represent a conservative estimate of the fragmentation effect as discussed in the text.
  }
\label{fig:ccfrag}
\end{figure}

To conclude this section, we would like to remark on a possible effect of fragmentation on the azimuthal correlations.  Fragmentation is traditionally taken into account by convoluting the parton spectra \eq{sincl},\eq{nlow} with the fragmentation functions $D(z)$. Particularly, hadron spectrum can be obtained from the gluon spectrum \eq{sincl} as following \cite{Kharzeev:2004yx,Tuchin:2007pf}
\beql{sincl-frag}
\frac{dN^h(k)}{d^2k dy}= \int_{z_\mathrm{min}}^1 \frac{dz}{z^2}\, \frac{dN^g(k/z)}{d^2k dy}\,D(z)\, F(k/z)\,,
\eeq
and similarly for the double-inclusive spectrum \eq{nlow}. Function $F(k)$ encodes  an effect of the DGLAP evolution at high $k$ \cite{Kharzeev:2004yx,Tuchin:2007pf}. We are required to include this function  because the integral in \eq{sincl-frag} extends to small values of $z$ up to the kinematically allowed $z_\mathrm{min}$. At small $z$'s  the high-$k$ tail of the parton spectrum, which is not accounted for in our model, plays an important role. Recall, that  in this paper we use an approximation by which integrals over the transverse momenta are approximated by the value of the integrand at the lower limit of integration (see the discussion above in this section). Therefore, we can approximate the $z$-integral in \eq{sincl-frag} be the value of the integrand at  $z_0\sim \mathcal{O}(1)$, where $z_0$ corresponds to the maximal value of the integrand as a function of $z$. The net result of this approximation  is a shift  of the parton spectrum towards higher momenta $k\to k/z_0$.
We checked that our  results shown in \fig{fig:cc} -- \fig{fig:ff2} are insensitive to such a shift in the region $1<z_0<1/3$. In order to set a limit on a maximal error that we make assuming this approximation, we performed an explicit integration over $z$ as follows. Using the experimental data on the hadron spectrum \cite{Adler:2003pb} we set  $F(k) \propto (1-k/k_0)^{-n}$, with $n=6$ and $k_0=1.2$~GeV. For the transverse momenta of interest here, this effectively leads to an additional factor of $z^6$ at small $z$ in the integrand of \eq{sincl-frag}. Fragmentation functions are taken from \cite{frag}. The corresponding modification of  azimuthal distributions  of \fig{fig:cc} is exhibited in \fig{fig:ccfrag}.
%%%%
%%%%%
We observe that even the most conservative estimate leads to a minor modification of the correlation functions in agreement with the approximations that we made. Similar conclusion holds also for the correlation functions at  forward rapidity.

%%%%%%%
\section{Correlations at $|y_T-y_A|\gg 1$}\label{sec3}

 If the trigger and associated particles are well separated in rapidity so that $|y_T-y_A|\sim 1/\as\gg 1$, we can apply the MRK approximation \eq{mrk} to the double-inclusive cross section. The result then factorizes into a product of two ladder rungs each given by  the real part of the LO BFKL kernel.  The corresponding formula is 
\beql{wincl}
\frac{dN_\mathrm{corr}}{d^2 k_T\,dy_T\, d^2 k_A\,dy_A}=
\frac{N_c \, \as^2}{\pi^2 \, C_F \, S_\bot}\,
\frac{1}{ k_T^2\, k_A^2}\,\int\,
d^2 q_1\,\varphi_D(x_1,q^2_1)\,\varphi_A(x_2,(\b k_T+\b k_A- \b q_1)^2)\,,
\eeq
where $x_{1,2}$ are defined in \eq{x12}. The advantage of the MRK approximation  is that it allows taking into account a possible multi-gluon production in the interval between $y_T$ and $y_A$. Unfortunately, in this approximation one also looses many  features of the azimuthal angle dependence that are important for  description of the backward  pick at  intermediate $\Delta y$, see \sec{sec2}. Thus, we are facing a dilemma: 
either to use formulas of \sec{sec2} that give precise dependence on  $\Delta y$ but neglect evolution in the gap, or to take into account the evolution as discussed below (see \eq{wev}-\eq{saddle}) but in the MRK limit $\Delta y\to \infty$. At present, there is no approach that would interpolate between these limits at intermediate $\Delta y$ relevant for RHIC. Therefore, in this section we will calculate the correlation function in two limits, compare our results with the data and try to learn which approximation is more phenomenologically important  at $\Delta y=3$.

Eq.~\eq{wincl} does not take into account a  possible gluon emission in the rapidity interval  between  $y_T$ and $y_A$. This is important  when $|y_1-y_2|> 1/\as$ and may be important for the experimentally measured forward-backward rapidity correlations. Evolution in between the  rapidities of the produced particles can be included using the 
 the AGK cutting rules \cite{Abramovsky:1973fm} and the known properties of the BFKL equation \cite{BFKL} as 
 \begin{eqnarray}\label{wev}
\frac{dN_\mathrm{corr}}{d^2 k_T\,dy_T\, d^2 k_A\,dy_A}&=&
\frac{N_c \, \as^2}{\pi^2 \, C_F \, S_\bot}\,
\frac{1}{ k_T^2\, k_A^2}\,\int
d^2 q_1\int
d^2 q_2\,\varphi_D(x_1,q^2_1)\,\varphi_A(x_2,(\b k_T+\b k_A- \b q_1)^2)\nonumber\\
&&\times \,G(|\b q_1-\b k_ T|, |\b q_2-\b k_A|, y_T-y_A)\,,
\end{eqnarray}
where $G$ is the Green's function of the BFKL equation \cite{BFKL}. It can be written as 
\beql{green}
G(q_1,q_2,y)=\sum_{n=0}^\infty 2\cos(n\,\unit q_1\cdot \unit q_2)\, G_n(q_1,q_2,y)\,.
\eeq
Functions $G _n$ are given by
\beq\label{fn}
G_n(q_1,q_2,y)=\frac{1}{2\pi^2 q_1 q_2}\int_{-\infty}^\infty d\nu \, \left(\frac{q_1}{q_2}\right)^{2i\nu}\,e^{2\bas \,\chi_n(\nu)\, y}\,,
\eeq
where 
\beql{chi}
\chi_n(\nu)=\psi(1)-\real\{\psi[(n+1)/2+i\nu]\}\,,
\eeq
and $\psi(\nu)= \Gamma'(z)/\Gamma(z)$.
One can evaluate the $\nu$-integral in \eq{fn} in the saddle point approximation, which is valid when $|\ln q_1/q_2|\ll \as y$. We have 
\beql{saddle}
G_n(q_1,q_2,y)\approx \frac{1}{2\pi^2 q_1 q_2}\,e^{2\bas \,\chi_n(0)\, y}\, \exp\bigg\{\frac{\ln^2(q_1/q_2)}{\bas \, y\, \psi_2[(n+1)/2]}\bigg\}\sqrt{\frac{2\pi}{-2\bas\,y\, \psi_2[(n+1)/2]}}\,,
\eeq
with $\psi_2(\nu)= \Gamma''(z)/\Gamma(z)$. The numerical values of poly-gamma functions for $n=1,2,3$ are listed below
\begin{subequations}\label{val}
\begin{eqnarray}
&&\chi_0(0)= \ln 4\,,\quad \chi_1(0)=0\,,\quad  \chi_2(0)=-2+\ln 4\,.\\
&&\psi_2(1/2)=-14\zeta(3)\,,\quad \psi_2(1)= -2\zeta(3)\,,\quad \psi_2(3/2)= -14\zeta(3)+16\,.
 \end{eqnarray}
\end{subequations}

It is easily seen that $G_0$ has positive intercept, $G_1$ has zero intercept and $G_2$, $G_3$, etc.\  have negative ones. Therefore, $G_0$ dominates at high $y$. However, $n=0$ term does not produce any azimuthal angle dependence. It is the sub-leading $n=1$ term that is responsible for the azimuthal correlations at large $y_T-y_A$. Thus, we use an approximation $G\approx 2[G_0+ G_1\cos(\unit q_1\cdot \unit q_2)]$ with $G_0$ and $G_1$ given by \eq{saddle}. This is the same approximation that was employed in \cite{Kharzeev:2004bw}. 
  
\begin{figure}[ht]
\begin{tabular}{cc}
      \includegraphics[height=4.5cm]{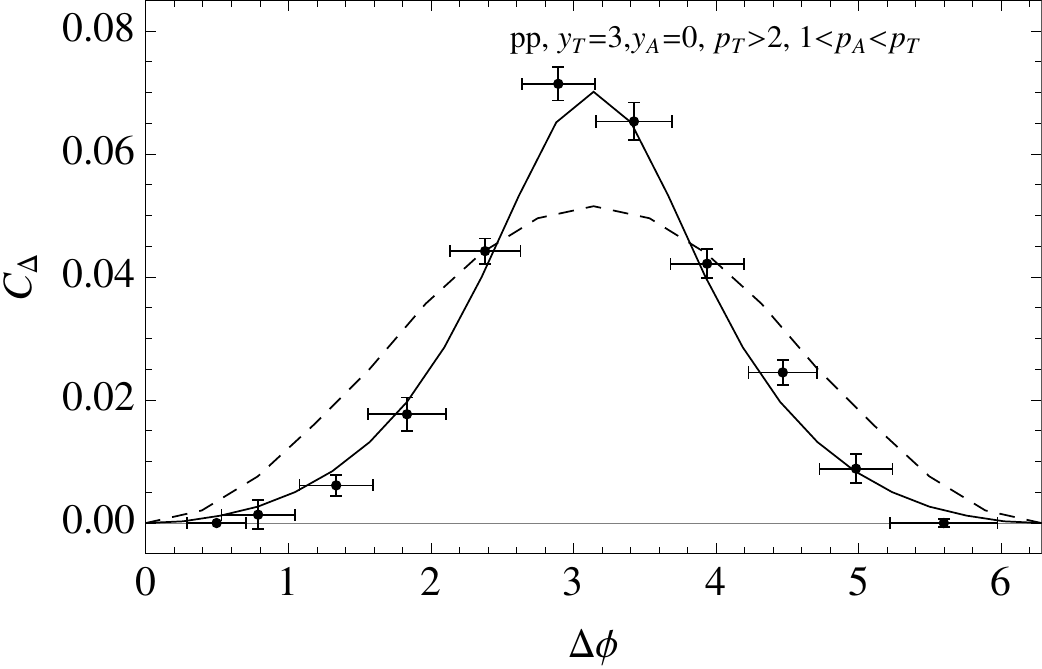} &
      \includegraphics[height=4.5cm]{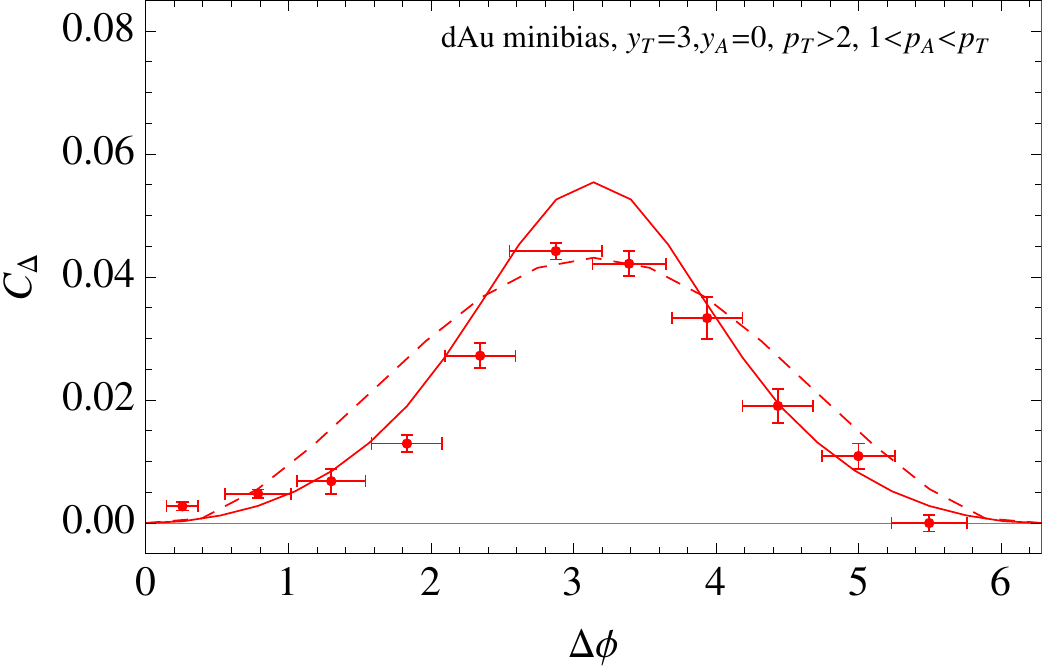} 
  \end{tabular}
  \caption{Forward-backward correlations. Kinematic region is  $p_T >2$, $1<p_A <p_T$ (all momenta are in GeV), $y_T=3$, $y_A=0$. Left (right) panel: minbias $pp$ ($dAu$) collisions. Solid lines: calculations with \eq{nlow} (exact $2\to 4$ amplitude, no evolution between the trigger and associate particles). Dashed line: calculations with \eq{wev} (MRK approximation of $2\to 4$ amplitude, includes evolution between  the trigger and associate particles).  Data  from \cite{Adams:2006uz,Braidot:2009ji} (forward $\pi^0$ and midrapidity $h^\pm$).}
\label{fig:fb}
\end{figure}

Numerical calculations are presented in \fig{fig:fb} together with the experimental data from \cite{Adams:2006uz,Braidot:2009ji}. We observe that the shape of the correlation function is better described by \eq{nlow}--\eq{ggqq} in agreement with the observation of \cite{Leonidov:1999nc,Kovchegov:2002cd} that  finite $\Delta y$ corrections to the MRK approximation are essential for description of azimuthal correlations. On the other hand, it seems that to explain the magnitude of depletion  one also needs to include the small $x$ evolution effects in the gap $\Delta y=y_T-y_A$. Obviously, a more accurate description requires additional theoretical investigation of the finite $\Delta y$ corrections.    Data  from \cite{Adams:2006uz,Braidot:2009ji} also shows  a significant isospin effect (not displayed here) that probably originates in the valence quark contribution not taken into account in the present work. This isospin effect obscures the CGC contribution and requires a detailed analyses that we plan to do elsewhere. At LHC one can get rid of the isospin effect by considering correlations at large rapidity gaps away from the fragmentation regions.

%%%%
%%%%%

%%%%%%%
\section{Conclusions}\label{sec:concl}

We calculated the azimuthal correlation function in $dAu$ collisions using the approach developed by us before \cite{Kovchegov:2002nf,Kovchegov:2002cd}. The results are presented in Figs.~\ref{fig:cc},\ref{fig:ff1},\ref{fig:ff2} and \ref{fig:fb}. We demonstrated that CGC is responsible for depletion of the back-to-back correlations in $dAu$ collisions as compared to those in $pp$ ones
at small rapidity separations -- at midrapidity and forward rapidity -- and at large rapidity separations.  Our results quantitatively confirm earlier arguments of \cite{Kharzeev:2004bw}.

%%%%%%%%%%%%%%%%%%%%%%%%%%%%%%%%
\acknowledgments

I am grateful to Larry McLerran for drawing my attention to Ref.~\cite{AGordon}, encouraging writing this paper and reading its draft version. I would like to thank him, Genya Levin and Cyrille Marquet   for useful discussions. 
This work  was supported in part by the U.S. Department of Energy under Grant No.\ DE-FG02-87ER40371. I 
thank RIKEN, BNL, and the U.S. Department of Energy (Contract No.\ DE-AC02-98CH10886) for providing facilities essential
for the completion of this work.

%%%%%%%%%%%%%%%%%%%%%%%%%%%%%%%%%%%%%


\begin{thebibliography}{80}

\bibitem{GLR}
L. V. Gribov, E. M. Levin and M. G. Ryskin, Phys. Rep. {\bf 100}
(1983) 1.
%%CITATION = PRPLC,100,1;%%

\bibitem{MUQI}
A. H. Mueller and J. Qiu,  Nucl. Phys. {\bf B 268} (1986) 427.
%%CITATION = NUPHA,B268,427;%%

\bibitem{MV}
L.~D.~McLerran and R.~Venugopalan,
%``Computing quark and gluon distribution functions for very large nuclei,''
Phys.\ Rev.\ D {\bf 49}, 2233 (1994)
[arXiv:hep-ph/9309289],
%%CITATION = HEP-PH 9309289;%%
%``Gluon distribution functions for very large nuclei at small transverse momentum,''
Phys.\ Rev.\ D {\bf 49}, 3352 (1994)
[arXiv:hep-ph/9311205],
%%CITATION = HEP-PH 9311205;%%
%``Green's functions in the color field of a large nucleus,''
Phys.\ Rev.\ D {\bf 50}, 2225 (1994)
[arXiv:hep-ph/9402335],
%%CITATION = HEP-PH 9402335;%%
%``Fock space distributions, structure functions, higher twists and  small x,''
Phys.\ Rev.\ D {\bf 59}, 094002 (1999)
[arXiv:hep-ph/9809427].
%%CITATION = HEP-PH 9809427;%%

 \bibitem{JIMWLK1}
J.~Jalilian-Marian, A.~Kovner, A.~Leonidov and H.~Weigert,
 Phys.\ Rev.\,  {\bf D59} (1999) 014014
[arXiv:hep-ph/9706377];\,\,  Nucl.\ Phys.\,{\bf B504} (1997) 415
[arXiv:hep-ph/9701284].

\bibitem{JIMWLK2}
E.~Iancu, A.~Leonidov and L.~D.~McLerran,
 Phys.\ Lett.\,  {\bf B510} (2001) 133
[arXiv:hep-ph/0102009];\,\,  Nucl.\ Phys.\,  {\bf A692} (2001) 583
[arXiv:hep-ph/0011241].

%\cite{Gyulassy:2004zy}
\bibitem{Gyulassy:2004zy}
  M.~Gyulassy and L.~McLerran,
  %``New forms of QCD matter discovered at RHIC,''
  Nucl.\ Phys.\  A {\bf 750}, 30 (2005)
  [arXiv:nucl-th/0405013].
  %%CITATION = NUPHA,A750,30;%%



%\cite{Kovchegov:2002nf}
\bibitem{Kovchegov:2002nf}
  Y.~V.~Kovchegov and K.~L.~Tuchin,
  %``Elliptic flow from minijet production in heavy ion collisions,''
  Nucl.\ Phys.\  A {\bf 708}, 413 (2002)
  [arXiv:hep-ph/0203213].
  %%CITATION = NUPHA,A708,413;%%

%\cite{Kovchegov:2002cd}
\bibitem{Kovchegov:2002cd}
  Y.~V.~Kovchegov and K.~L.~Tuchin,
  %``Correlation functions and cumulants in elliptic flow analysis,''
  Nucl.\ Phys.\  A {\bf 717}, 249 (2003)
  [arXiv:nucl-th/0207037].
  %%CITATION = NUPHA,A717,249;%%

%\cite{Tang:2004vc}
\bibitem{Tang:2004vc}
  A.~H.~Tang  [STAR Collaboration],
  %``Azimuthal anisotropy and correlations in p + p, d + Au and Au + Au
  %collisions at 200-GeV,''
  J.\ Phys.\ G {\bf 30}, S1235 (2004)
  [arXiv:nucl-ex/0403018].
  %%CITATION = JPHGB,G30,S1235;%%

%\cite{Kharzeev:2004bw}
\bibitem{Kharzeev:2004bw}
  D.~Kharzeev, E.~Levin and L.~McLerran,
  %``Jet azimuthal correlations and parton saturation in the color glass
  %condensate,''
  Nucl.\ Phys.\  A {\bf 748}, 627 (2005)
  [arXiv:hep-ph/0403271].
  %%CITATION = NUPHA,A748,627;%%

%\cite{Kovchegov:1998bi}
\bibitem{Kovchegov:1998bi}
  Y.~V.~Kovchegov and A.~H.~Mueller,
  %``Gluon production in current nucleus and nucleon nucleus collisions in  a
  %quasi-classical approximation,''
  Nucl.\ Phys.\  B {\bf 529}, 451 (1998)
  [arXiv:hep-ph/9802440].
  %%CITATION = NUPHA,B529,451;%%


%\cite{Levin:1999mw}
\bibitem{Levin:1999mw}
  E.~Levin and K.~Tuchin,
  %``Solution to the evolution equation for high parton density QCD,''
  Nucl.\ Phys.\  B {\bf 573}, 833 (2000)
  [arXiv:hep-ph/9908317].
  %%CITATION = NUPHA,B573,833;%%

%\cite{Kharzeev:2004yx}
\bibitem{Kharzeev:2004yx}
  D.~Kharzeev, Y.~V.~Kovchegov and K.~Tuchin,
  %``Nuclear modification factor in d + Au collisions: Onset of suppression  in
  %the color glass condensate,''
  Phys.\ Lett.\  B {\bf 599}, 23 (2004)
  [arXiv:hep-ph/0405045].
  %%CITATION = PHLTA,B599,23;%%

%\cite{Albacete:2003iq}
\bibitem{Albacete:2003iq}
  J.~L.~Albacete, N.~Armesto, A.~Kovner, C.~A.~Salgado and U.~A.~Wiedemann,
  %``Energy dependence of the Cronin effect from non-linear QCD evolution,''
  Phys.\ Rev.\ Lett.\  {\bf 92}, 082001 (2004)
  [arXiv:hep-ph/0307179].
  %%CITATION = PRLTA,92,082001;%%



%\cite{Kharzeev:2003wz}
\bibitem{Kharzeev:2003wz}
  D.~Kharzeev, Y.~V.~Kovchegov and K.~Tuchin,
  %``Cronin effect and high-p(T) suppression in p A collisions,''
  Phys.\ Rev.\  D {\bf 68}, 094013 (2003)
  [arXiv:hep-ph/0307037].
  %%CITATION = PHRVA,D68,094013;%%

\bibitem{dipole}
A.~H.~Mueller,
%``Small X Behavior And Parton Saturation: A QCD Model,''
Nucl.\ Phys.\ B {\bf 335}, 115 (1990);
%%CITATION = NUPHA,B335,115;%%
%\cite{Mueller:1993rr}
%A.~H.~Mueller,
%``Soft gluons in the infinite momentum wave function and the BFKL pomeron,''
Nucl.\ Phys.\ B {\bf 415}, 373 (1994);
%%CITATION = NUPHA,B415,373;%%
%\cite{Mueller:1994jq}
A.~H.~Mueller and B.~Patel,
%``Single and double BFKL pomeron exchange and a dipole picture of high-energy hard processes,%''
Nucl.\ Phys.\ B {\bf 425}, 471 (1994)
[arXiv:hep-ph/9403256];
%%CITATION = HEP-PH 9403256;%%

%\cite{JalilianMarian:2004da}
\bibitem{JalilianMarian:2004da}
  J.~Jalilian-Marian and Y.~V.~Kovchegov,
  %``Inclusive two-gluon and valence quark-gluon production in DIS and p A,''
  Phys.\ Rev.\  D {\bf 70}, 114017 (2004)
  [Erratum-ibid.\  D {\bf 71}, 079901 (2005)]
  [arXiv:hep-ph/0405266].
  %%CITATION = PHRVA,D70,114017;%%

%\cite{Tuchin:2004rb}
\bibitem{Tuchin:2004rb}
  K.~Tuchin,
  %``Heavy quark production by a quasi-classical color field in proton  nucleus
  %collisions,''
  Phys.\ Lett.\  B {\bf 593}, 66 (2004)
  [arXiv:hep-ph/0401022].
  %%CITATION = PHLTA,B593,66;%%

%\cite{Blaizot:2004wv}
\bibitem{Blaizot:2004wv}
  J.~P.~Blaizot, F.~Gelis and R.~Venugopalan,
  %``High energy p A collisions in the color glass condensate approach. II:
  %Quark production,''
  Nucl.\ Phys.\  A {\bf 743}, 57 (2004)
  [arXiv:hep-ph/0402257].
  %%CITATION = NUPHA,A743,57;%%


%\cite{Kovchegov:2006qn}
\bibitem{Kovchegov:2006qn}
  Y.~V.~Kovchegov and K.~Tuchin,
  %``Production of q anti-q pairs in proton nucleus collisions at high
  %energies,''
  Phys.\ Rev.\  D {\bf 74}, 054014 (2006)
  [arXiv:hep-ph/0603055].
  %%CITATION = PHRVA,D74,054014;%%


\bibitem{Marquet:2007vb}
  C.~Marquet,
  %``Forward inclusive dijet production and azimuthal correlations in pA
  %collisions,''
  Nucl.\ Phys.\  A {\bf 796}, 41 (2007)
  [arXiv:0708.0231 [hep-ph]].


%\cite{Leonidov:1999nc}
\bibitem{Leonidov:1999nc}
  A.~Leonidov and D.~Ostrovsky,
  %``Angular and momentum asymmetry in particle production at high-energies,''
  Phys.\ Rev.\  D {\bf 62}, 094009 (2000)
  [arXiv:hep-ph/9905496].
  %%CITATION = PHRVA,D62,094009;%%
  
 %\cite{Bartels:2006hg}
\bibitem{Bartels:2006hg}
  J.~Bartels, A.~Sabio Vera and F.~Schwennsen,
  %``NLO inclusive jet production in k(T)-factorization,''
  JHEP {\bf 0611}, 051 (2006)
  [arXiv:hep-ph/0608154].
  %%CITATION = JHEPA,0611,051;%%


 \bibitem{LRSS}
E.~M.~Levin, M.~G.~Ryskin, Y.~M.~Shabelski and A.~G.~Shuvaev,
%``Heavy quark production in semihard nucleon interactions,''
Sov.\ J.\ Nucl.\ Phys.\  {\bf 53}, 657 (1991)
[Yad.\ Fiz.\  {\bf 53}, 1059 (1991)].
%%CITATION = SJNCA,53,657;%%

\bibitem{CCH}
S.~Catani, M.~Ciafaloni and F.~Hautmann,
%``High-Energy Factorization And Small X Heavy Flavor Production,''
Nucl.\ Phys.\ B {\bf 366}, 135 (1991).
%%CITATION = NUPHA,B366,135;%%

\bibitem{CE} 
J.~C.~Collins and R.~K.~Ellis,
Nucl.\ Phys.\ B {\bf 360}, 3 (1991).
%%CITATION = NUPHA,B360,3;%% 

%\cite{Fadin:1996zv}
\bibitem{Fadin:1996zv}
  V.~S.~Fadin, M.~I.~Kotsky and L.~N.~Lipatov,
  %``Gluon pair production in the quasi-multi-Regge kinematics,''
  arXiv:hep-ph/9704267.
  %%CITATION = HEP-PH/9704267;%%

%\cite{Fadin:1997hr}
\bibitem{Fadin:1997hr}
  V.~S.~Fadin, R.~Fiore, A.~Flachi and M.~I.~Kotsky,
  %``Quark-antiquark contribution to the BFKL kernel,''
  Phys.\ Lett.\  B {\bf 422}, 287 (1998)
  [arXiv:hep-ph/9711427].
  %%CITATION = PHLTA,B422,287;%%


%\cite{Kovchegov:2001sc}
\bibitem{Kovchegov:2001sc}
  Y.~V.~Kovchegov and K.~Tuchin,
  %``Inclusive gluon production in DIS at high parton density,''
  Phys.\ Rev.\  D {\bf 65}, 074026 (2002)
  [arXiv:hep-ph/0111362].
  %%CITATION = PHRVA,D65,074026;%%

%\cite{Fujii:2005vj}
\bibitem{Fujii:2005vj}
  H.~Fujii, F.~Gelis and R.~Venugopalan,
  %``Quantitative study of the violation of k(T)-factorization in
  %hadroproduction of quarks at collider energies,''
  Phys.\ Rev.\ Lett.\  {\bf 95}, 162002 (2005)
  [arXiv:hep-ph/0504047].
  %%CITATION = PRLTA,95,162002;%%

%\cite{Kharzeev:2000ph}
\bibitem{Kharzeev:2000ph}
  D.~Kharzeev and M.~Nardi,
  %``Hadron production in nuclear collisions at RHIC and high density QCD,''
  Phys.\ Lett.\  B {\bf 507}, 121 (2001)
  [arXiv:nucl-th/0012025].
  %%CITATION = PHLTA,B507,121;%%

%\cite{Kharzeev:2001yq}
\bibitem{Kharzeev:2001yq}
  D.~Kharzeev, E.~Levin and M.~Nardi,
  %``The onset of classical QCD dynamics in relativistic heavy ion
  %collisions,''
  Phys.\ Rev.\  C {\bf 71}, 054903 (2005)
  [arXiv:hep-ph/0111315].
  %%CITATION = PHRVA,C71,054903;%%

%\cite{Kharzeev:2001gp}
\bibitem{Kharzeev:2001gp}
  D.~Kharzeev and E.~Levin,
  %``Manifestations of high density QCD in the first RHIC data,''
  Phys.\ Lett.\  B {\bf 523}, 79 (2001)
  [arXiv:nucl-th/0108006].
  %%CITATION = PHLTA,B523,79;%%
 
 %\cite{Kharzeev:2002ei}
\bibitem{Kharzeev:2002ei}
  D.~Kharzeev, E.~Levin and M.~Nardi,
  %``QCD saturation and deuteron nucleus collisions,''
  Nucl.\ Phys.\  A {\bf 730}, 448 (2004)
  [Erratum-ibid.\  A {\bf 743}, 329 (2004)]
  [arXiv:hep-ph/0212316].
  %%CITATION = NUPHA,A730,448;%%
 
  %\cite{Dumitru:2008wn}
\bibitem{Dumitru:2008wn}
  A.~Dumitru, F.~Gelis, L.~McLerran and R.~Venugopalan,
  %``Glasma flux tubes and the near side ridge phenomenon at RHIC,''
  Nucl.\ Phys.\  A {\bf 810}, 91 (2008)
  [arXiv:0804.3858 [hep-ph]].
  %%CITATION = NUPHA,A810,91;%%
  
  %\cite{Gelis:2008ad}
\bibitem{Gelis:2008ad}
  F.~Gelis, T.~Lappi and R.~Venugopalan,
  %``High energy factorization in nucleus-nucleus collisions II - Multigluon
  %correlations,''
  Phys.\ Rev.\  D {\bf 78}, 054020 (2008)
  [arXiv:0807.1306 [hep-ph]].
  %%CITATION = PHRVA,D78,054020;%%

%\cite{Gelis:2008sz}
\bibitem{Gelis:2008sz}
  F.~Gelis, T.~Lappi and R.~Venugopalan,
  %``High energy factorization in nucleus-nucleus collisions. 3. Long range
  %rapidity correlations,''
  Phys.\ Rev.\  D {\bf 79}, 094017 (2009)
  [arXiv:0810.4829 [hep-ph]].
  %%CITATION = PHRVA,D79,094017;%%

%\cite{Dusling:2009ni}
\bibitem{Dusling:2009ni}
  K.~Dusling, F.~Gelis, T.~Lappi and R.~Venugopalan,
  %``Long range two-particle rapidity correlations in A+A collisions from high
  %energy QCD evolution,''
  arXiv:0911.2720 [hep-ph].
  %%CITATION = ARXIV:0911.2720;%%

%\cite{Levin:1974fw}
\bibitem{Levin:1974fw}
  E.~M.~Levin and M.~G.~Ryskin,
  %``Inclusive Sections Of The Particle Production With Large Transverse
  %Impulses In Multiperipheral And Parton Models,''
  Yad.\ Fiz.\  {\bf 21}, 1072 (1975).
  %%CITATION = YAFIA,21,1072;%%

%\cite{Laenen:1994gh}
\bibitem{Laenen:1994gh}
  E.~Laenen and E.~Levin,
  %``Parton densities at high-energy,''
  Ann.\ Rev.\ Nucl.\ Part.\ Sci.\  {\bf 44}, 199 (1994).
  %%CITATION = ARNUA,44,199;%%

%\cite{Kovchegov:1997ke}
\bibitem{Kovchegov:1997ke}
  Y.~V.~Kovchegov and D.~H.~Rischke,
  %``Classical gluon radiation in ultrarelativistic nucleus nucleus
  %collisions,''
  Phys.\ Rev.\  C {\bf 56}, 1084 (1997)
  [arXiv:hep-ph/9704201].
  %%CITATION = PHRVA,C56,1084;%%

%\cite{Gyulassy:1997vt}
\bibitem{Gyulassy:1997vt}
  M.~Gyulassy and L.~D.~McLerran,
  %``Yang-Mills radiation in ultrarelativistic nuclear collisions,''
  Phys.\ Rev.\  C {\bf 56}, 2219 (1997)
  [arXiv:nucl-th/9704034].
  %%CITATION = PHRVA,C56,2219;%%


%\cite{Braun:2000ca}
\bibitem{Braun:2000ca}
  M.~A.~Braun,
  %``Comments on parton saturation at small x in large nuclei,''
  arXiv:hep-ph/0010041.
  %%CITATION = HEP-PH/0010041;%%

%\cite{Braun:2001kh}
\bibitem{Braun:2001kh}
  M.~A.~Braun,
  %``High-energy interaction with the nucleus in the perturbative QCD with  N(c)
  %--> infinity,''
  arXiv:hep-ph/0101070.
  %%CITATION = HEP-PH/0101070;%%



%\cite{Blaizot:2004wu}
\bibitem{Blaizot:2004wu}
  J.~P.~Blaizot, F.~Gelis and R.~Venugopalan,
  %``High energy p A collisions in the color glass condensate approach. I:
  %Gluon production and the Cronin effect,''
  Nucl.\ Phys.\  A {\bf 743}, 13 (2004)
  [arXiv:hep-ph/0402256].
  %%CITATION = NUPHA,A743,13;%%
  
  
%\cite{GolecBiernat:1999qd}
\bibitem{GolecBiernat:1999qd}
  K.~J.~Golec-Biernat and M.~Wusthoff,
  %``Saturation in diffractive deep inelastic scattering,''
  Phys.\ Rev.\  D {\bf 60}, 114023 (1999)
  [arXiv:hep-ph/9903358].
  %%CITATION = PHRVA,D60,114023;%%
  
  %\cite{GolecBiernat:1998js}
\bibitem{GolecBiernat:1998js}
  K.~J.~Golec-Biernat and M.~Wusthoff,
  %``Saturation effects in deep inelastic scattering at low Q**2 and its
  %implications on diffraction,''
  Phys.\ Rev.\  D {\bf 59}, 014017 (1999)
  [arXiv:hep-ph/9807513].
  %%CITATION = PHRVA,D59,014017;%%
  
 %\cite{Adams:2003im}
\bibitem{Adams:2003im}
  J.~Adams {\it et al.}  [STAR Collaboration],
  %``Evidence from d + Au measurements for final-state suppression of high  p(T)
  %hadrons in Au + Au collisions at RHIC,''
  Phys.\ Rev.\ Lett.\  {\bf 91}, 072304 (2003)
  [arXiv:nucl-ex/0306024].
  %%CITATION = PRLTA,91,072304;%%

\bibitem{AGordon}
A.~Gordon (for the STAR Collaboration), Presentation at the 3rd Joint Meeting of APS Division of Nuclear Physics and Physical Society of Japan, Hawaii, October 13-17, 2009.

%\cite{Nikolaev:2005dd}
\bibitem{Nikolaev:2005dd}
  N.~N.~Nikolaev, W.~Schafer, B.~G.~Zakharov and V.~R.~Zoller,
  %``Nonlinear k(T)-factorization for quark-gluon dijet production off
  %nuclei,''
  Phys.\ Rev.\  D {\bf 72}, 034033 (2005)
  [arXiv:hep-ph/0504057].
  %%CITATION = PHRVA,D72,034033;%%

%\cite{Itakura:2003jp}
\bibitem{Itakura:2003jp}
  K.~Itakura, Y.~V.~Kovchegov, L.~McLerran and D.~Teaney,
  %``Baryon stopping and valence quark distribution at small x,''
  Nucl.\ Phys.\  A {\bf 730}, 160 (2004)
  [arXiv:hep-ph/0305332].
  %%CITATION = NUPHA,A730,160;%%


%\cite{Tuchin:2007pf}
\bibitem{Tuchin:2007pf}
  K.~Tuchin,
  %``Forward hadron production in high energy p A collisions: From RHIC to
  %LHC,''
  Nucl.\ Phys.\  A {\bf 798}, 61 (2008)
  [arXiv:0705.2193 [hep-ph]].
  %%CITATION = NUPHA,A798,61;%%

%\cite{Adler:2003pb}
\bibitem{Adler:2003pb}
  S.~S.~Adler {\it et al.}  [PHENIX Collaboration],
  %``Mid-rapidity neutral pion production in proton proton collisions at
  %$\sqrt{s}$ = 200-GeV,''
  Phys.\ Rev.\ Lett.\  {\bf 91}, 241803 (2003)
  [arXiv:hep-ex/0304038].
  %%CITATION = PRLTA,91,241803;%%

\bibitem{frag}
B.~A.~Kniehl, G.~Kramer and B.~Potter,
Nucl.\ Phys.\ B {\bf 597}, 337 (2001)
[arXiv:hep-ph/0011155].
%%CITATION = HEP-PH 0011155;%%

%\cite{Abramovsky:1973fm}
\bibitem{Abramovsky:1973fm}
  V.~A.~Abramovsky, V.~N.~Gribov and O.~V.~Kancheli,
  %``CHARACTER OF INCLUSIVE SPECTRA AND FLUCTUATIONS PRODUCED IN INELASTIC
  %PROCESSES BY MULTI - POMERON EXCHANGE,''
  Yad.\ Fiz.\  {\bf 18}, 595 (1973)
  [Sov.\ J.\ Nucl.\ Phys.\  {\bf 18}, 308 (1974)].
  %%CITATION = SJNCA,18,308;%%


\bibitem{BFKL}
E.A. Kuraev, L.N. Lipatov and V.S. Fadin, {\em Sov. Phys. JETP} {\bf
45}, 199 (1977); Ya.Ya. Balitsky and L.N. Lipatov, {\em Sov. J. Nucl.
Phys.} {\bf 28}, 22 (1978).

%\cite{Adams:2006uz}
\bibitem{Adams:2006uz}
  J.~Adams {\it et al.}  [STAR Collaboration],
  %``Forward neutral pion production in p+p and d+Au collisions at  s(NN)**(1/2)
  %= 200-GeV,''
  Phys.\ Rev.\ Lett.\  {\bf 97}, 152302 (2006)
  [arXiv:nucl-ex/0602011].
  %%CITATION = PRLTA,97,152302;%%


%\cite{Braidot:2009ji}
\bibitem{Braidot:2009ji}
  E.~Braidot  [STAR collaboration],
  %``Looking forward for Color Glass Condensate signatures,''
  Nucl.\ Phys.\  A {\bf 830}, 603C (2009)
  [arXiv:0907.3473 [nucl-ex]].
  %%CITATION = NUPHA,A830,603C;%%













  
  
  
\end{thebibliography}
\end{document}